\documentstyle[psfig,floats,epsf,prl,aps,pstricks,psfig,pst-plot,rotate]{revtex}

\begin{document}

\twocolumn[\hsize\textwidth\columnwidth\hsize\csname
@twocolumnfalse\endcsname

\title{The Effect of the Environment on $\alpha$-Al$_{2}$O$_{3}$ (0001)
Surface Structures}
\author{Xiao-Gang Wang$^{1}$, Anne Chaka$^{1, 2}$,
Matthias Scheffler$^{1}$}
\address{$^{1}$Fritz-Haber-Institut
der Max-Planck-Gesellschaft, Faradayweg 4-6,
D-14195 Berlin-Dahlem, Germany\\
$^{2}$The Lubrizol Corporation,
29400 Lakeland Blvd., Wickliffe, OH  44092-2298, USA.}
\date{\today}
\maketitle

\begin{abstract}
We report that calculating the Gibbs free energy of the $\alpha$-Al$_2$O$_3$
(0001) surfaces in equilibrium with a realistic environment containing both
oxygen and hydrogen species is
essential for obtaining theoretical predictions consistent
with experimental observations.
Using density-functional theory
we find that even under conditions of
high oxygen partial pressure, the metal terminated surface is 
surprisingly stable. An oxygen terminated $\alpha$-Al$_2$O$_3$ (0001)
surface becomes stable only if hydrogen is present on the surface.
In addition, including hydrogen on the surface resolves discrepancies
between previous theoretical work and experimental results with respect
to the magnitude and direction of surface relaxations.
\vspace{0.2cm}


\end{abstract}
\vskip2pc]

The nature of the corundum surface ($\alpha$-Al$_2$O$_3$) is of
considerable importance in a wide variety of technological
applications. These range from catalytic supports and thin-film
substrates to corrosion and wear protection in mechanical systems.
Yet, despite considerable experimental and
theoretical efforts over the years\cite{VEH94}, the surface structure,
and even the surface stoichiometry, is a matter of strong controversy.
The principal reason for this is that it is
difficult to prepare clean, uniform surfaces with specific,
well-defined structures and stoichiometries.
The typical heterogeneity of the surface makes the
interpretation of low energy electron
diffraction (LEED) and other surface spectroscopic data difficult.
In addition, corundum is an insulator which makes the application of
STM and other techniques based on electron spectroscopy problematic.

Knowing the structure and stoichiometry of the corundum surface, however, is essential
for understanding the electronic, mechanical, and chemical properties which determine
its reactivity and performance in various applications.
Yet it has only been within the past two years that
new experimental techniques have enabled the chemical identification of the
surface terminations for these systems. Renaud has reported a
$(1 \times 1)$
structure, prepared by heating in an oxygen-rich atmosphere, which
could reasonably be interpreted as being O-terminated \cite{GR98}. Toofan and
Watson found in a tensor LEED experiment 
both Al- and O-terminated domains in a 2:1 ratio, respectively
\cite{JT98}. Ahn and Rabelais annealed the surface under UHV conditions
and determined the detailed structure of the Al-terminated surface
using time-of-flight scattering and recoiling spectrometry (TOF-SARS)\cite{JA97}.
The sensitivity of TOF-SARS also enabled the detection of
hydrogen randomly distributed on the surface which was stable even at
an annealing temperature of 1100$^\circ$~C.

Despite recent advances in experimental techniques, however, many questions
and inconsistencies remain that theoretical calculations performed to date
have not been able to resolve. 
Corundum ($\alpha$-Al$_{2}$O$_{3}$) crystallizes in a structure which
can be described by a primitive rhombohedral unit cell
with two Al$_{2}$O$_{3}$ formula units (10 atoms)
or by a conventional hexagonal unit cell
with six Al$_{2}$O$_{3}$ formula units (30 atoms). For the
hexagonal unit cell, the atoms are
stacked along the (0001) direction according to the sequence
$R$-AlAlO$_{3}$-AlAlO$_{3}$-$R$, where $R$ represents the 
continuing sequence in the bulk. The oxygen atoms form
hcp layers, and the metal atoms fill two thirds
of the octahedral sites between these layers\cite{RWGW82}. 

In the $R$-Al-Al-O$_{3}$-$R$ corundum 
structure (0001) stacking sequence, there are three unique
stoichiometric slice planes.
Yet theoretical methods ranging from the empirical to the
{\it ab initio}\cite{MAC87,CAU89,GUO92,IM93,GOD94,CV98} have so far
identified only one stable termination stoichiometry for the
$\alpha$-Al$_{2}$O$_{3}$ (0001) surface:  an Al-monolayer indicated by
AlO${_3}$Al-$R$ in Fig. 1. No stable O-terminated surface has
been found, so how can the experimentally observed
O-terminations be explained? In addition to the surface
stoichiometry, there is considerable disagreement with respect to the
magnitude and direction of the surface relaxations. For the 
Al-terminated surface, experimentally observed relaxations range from
-51\%\cite{PG97} to -63\%\cite{JA97}, whereas the
density-functional theory (DFT) pseudopotential 
calculations predict a relaxation
of about -87\%\cite{IM93,CV98}, and Hartree-Fock gives -40\%\cite{CAU89}. 
Toofan and Watson reported an outward relaxation for their
two domain system,
which is the opposite of what other experimental and theoretical
investigations had concluded. These discrepancies are significant
and need to be resolved.

What has not been previously considered in the theoretical treatment of the
$\alpha$-Al$_{2}$O$_{3}$ (0001) surface is the effect of the environment on
the surface structures and stoichiometry. Under realistic conditions, a
surface will exchange atoms with its surroundings. Hence in this work we
present the first analysis of the Gibbs free energy of the system with
respect to its dependence on the chemical potentials of the components present in
the material and the environment at 0K and 1000K. For a metal oxide such as
$\alpha$-Al$_{2}$O$_{3}$, the O$_{2}$ partial pressure is obviously the most
important factor in the analysis, as well as temperature. In addition, 
the presence of stable hydrogen on
the surface also needs to be addressed, as it can be incorporated into the
bulk structure during growth, remain from calcination of Al(OH)$_3$ 
in the synthesis of $\alpha$-Al$_{2}$O$_{3}$, 
or may result from exposure of the surface to
water vapor prior to placement in the UHV chamber. 
Yet to our knowledge the
presence of hydrogen on the corundum surface has not yet been investigated
in theoretical studies. In this paper we report that the surface
stoichiometries, structures, and properties change significantly depending
upon the chemical potential of O$_{2}$, H$_{2}$, and H$_{2}$O gases in
equilibrium with the surface at different temperatures. 
We use the full-potential linearized augmented
planewave (FP-LAPW) method to
solve the Kohn-Sham equations and calculate the total energies, forces, and
chemical potentials for all reasonable $(1 \times 1)$ corundum (0001) surface
geometries. We
find that the metal-terminated surface is surprisingly stable across the
range of a physically realistic oxygen chemical potential.
An oxygen terminated surface becomes
stable only if hydrogen is present on the surface, even at partial O$_{2}$
pressures which are too high for standard UHV equipment. In addition,
including hydrogen on the surface of both the aluminum and oxygen terminated
surfaces results in calculated relaxations which agree with the latest
experimental results, in both magnitude and direction.

\begin{figure}[t]
 \pspicture(5,5.0)
 \rput[c](4.3,2.5){
 \psfig{figure=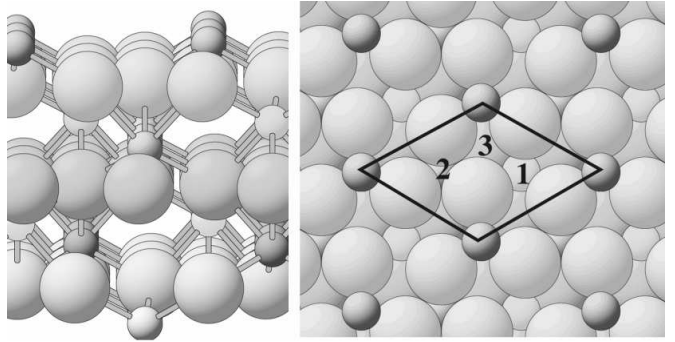,height=4.5cm}}
 \endpspicture
\caption{
Side and top view of the Al-O${_3}$-Al-$R$ surface of $\alpha$-Al$_2$O$_3$ ,
in which 1, 2 and 3 indicate three sites which could be occupied by the
topmost Al instead of 
the bulk site. The large circles denote O atoms and the small
circles Al atoms.
}
\label{Al-surf}
\end{figure}

To systematically investigate the $(1 \times 1)$ (0001) surface which is
observed at annealing temperatures below 1250$^\circ$~C, we have generated
what we believe are all the possible ($1 \times 1$) (0001) Al- and/or
O-surface terminations of Al$_2$O$_3$, plus several
additional surfaces containing hydroxyl groups.
Previous studies which investigated possible ($1 \times 1$) (0001)
surface terminations examined only the three types of structures
obtained by simply cleaving the unit cell at unique
positions\cite{GUO92,GOD94}: 
AlO${_3}$Al-$R$, O${_3}$AlAl-$R$, and AlAlO${_3}$-$R$. 
For the single Al-terminated layer there are three other possible 
locations for the Al atom in addition to the corresponding bulk site,
as shown in Fig.~\ref{Al-surf}.
For the O-terminated surface, other investigations have not considered
the possibility of oxygen vacancies which could occur and still 
maintain a perfect ($1 \times 1$) surface periodicity.
Two additional O-terminated structures can be created by introducing
one or two oxygen vacancies per unit cell, indicated by
O${_2}$AlAl-$R$ and O${_1}$AlAl-$R$, respectively.
Hence, there are a total of at least eight possible ($1 \times 1 $)
geometries of the (0001) surface which we investigate to determine
their relative stability.

In our calculations, the $(1 \times 1)$ $\alpha$-Al$_2$O$_3$ surface 
is modeled by a slab which consists of a finite number of layers 
and is infinite in the plane of the surface.
The slabs are repeated periodically along the [0001] direction 
and separated by 10\AA~ of vacuum.
The slab contains six oxygen O$_3$ layers and from ten to fourteen aluminum 
layers, depending upon the specific surface studied. 
We carefully tested
that the thickness of the vacuum as well as that of the slabs are sufficiently
large to ensure that surface-surface interactions through both the vacuum
and the slab are negligible.\cite{WCS00} The two surfaces of the slab are
identical and inversion symmetry is maintained. 

For our total-energy calculations we use the generalized gradient
approximation of Perdew {\em et al.}\cite{JPP92}
for the exchange-correlation potential
and the FP-LAPW method as implemented within the
WIEN97 program \cite{PB90,BK96,MP99}
to solve the Kohn-Sham equations\cite{parameters}.
A uniform {\bf k}-point mesh with ten points
is used for the entire surface Brillouin zone.
An identical mesh is used for the bulk calculations
to ensure consistency.
The calculated equilibrium lattice constants for the hexagonal
unit cell and the non-symmetry fixed positions of the atoms agree with
experimental values to within 0.7\%. 
The calculations give the heat of formation at 0K of the bulk Al$_{2}$O$_{3}$,
$\Delta{H\rm _{f}^{o}}$=17.37 eV, which is in good agreement with
the experimental value of 17.24 eV \cite{CRC87}. 

To compare the stability of surfaces with different stoichiometries
in chemical and thermal equilibrium with the gas phase and with the bulk, we
calculate the Gibbs free energy $\Omega$ of the
surface relative to the chemical potential of oxygen. The Gibbs
free surface energy $\Omega$ of a slab at temperature $T$ and partial
pressure $p$ is given by
$\Omega = E^{\rm total} + \Delta G^{\rm vib} - \sum N_{i}\mu_{i}(T,p)$,
where $E^{\rm total}$ is the scf energy of the slab,
$\Delta G^{\rm vib}$ is the vibrational contribution to the Gibbs
free energy,
$N_{i}$ the number of $i$th type of atoms in the slab, and 
$\mu_{i}(T,p)$ is the chemical potential
of the $i$th type of atom at a given temperature and pressure.
The sum is over all the types of the atoms in the slab.
To calculate the values of the chemical potential for all
species we use $\mu_{i}(T,p)$ = $\mu_{i}^{0}$ + 
$\Delta\mu_{i}(T,p)$, where 0K is taken as
the reference state, and $\Delta\mu_{i}(T,p)$ is the change
in free energy from that reference state to the system at a given temperature
and pressure. The 0K values for surfaces and for bulk
Al and Al$_{2}$O$_{3}$ are taken from our total energy
calculations. For the dissociation energies of the 
H$_2$, O$_2$, and H$_2$O molecules,
we use experimental values.
The $\Delta\mu_{i}(T,p)$
values for all crystalline and gas phase species are taken from
the JANAF Thermochemical Tables \cite{JANAF}. 

\begin{figure}
\psfig{figure=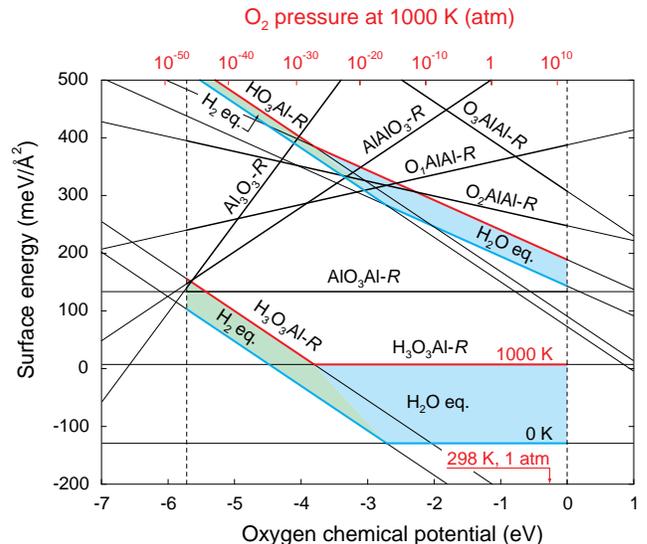,width=8.5cm}
\caption{Surface energies of different Al$_2$O$_3$(0001) surface
terminations. The dashed verticle lines indicate the allowed range of 
the oxygen chemical potential, $\mu_{\rm O}$. Values in red indicate
examples of higher temperatures and pressures. 
The green and blue regions indicate the range where
hydrogen on the surface is in equilibrium with 
H$_2$ and H$_2$O, respectively, from 0K up to 1000K and 1 atm
pressure (red lines). 
}
\label{E-surf}
\end{figure}

As each surface is considered here to be in chemical and thermal equilibrium
with the bulk and the environment, the chemical potentials
are constrained, so for the clean surface:
$2\mu_{\rm Al}+3\mu_{\rm O}=E_{\rm Al_{2}O_{3}}^{\rm bulk}$,
$\mu_{\rm Al} < E_{\rm Al}^{\rm bulk}$, and
$\mu_{\rm O} < \frac{1}{2}E_{{\rm O}_2}^{\rm molecule}$, 
where E$_{{\rm Al}_2{\rm O}_3}^{\rm bulk}$ is the total energy per bulk
Al$_2$O$_3$ formula unit.
The limits of the chemical potential for oxygen are determined by
conditions of equilibrium with relevant oxygen-containing species
in the system.  The maximum oxygen chemical potential is
that of a maximum concentration of O$_2$ 
molecules at 0K, which
corresponds to O$_2$ 
condensing on the surface.  This is the reference value of
the oxygen chemical potential, which is the zero on the right hand
side of the graph in Fig. \ref{E-surf}.
Since the system is in equilibrium with bulk
Al$_2$O$_3$ the minimum $\mu_{\rm O}$ occurs when
$\mu_{\rm Al}$ is at a maximum and 
$\mu_{\rm O}=\frac{1}{3}E_{\rm Al_{2}O_{3}}^{\rm bulk}-
\frac{2}{3}E_{\rm Al}^{\rm bulk}$. 
Below this range, aluminum metal would condense on the surface.

For the surfaces where hydrogen is adsorbed, the free energy of the surface
is calculated relative to the chemical potential
of hydrogen $\mu_{\rm H}$ in both H$_2$ and H$_2$O,
as desorption of hydrogen as either molecular species is possible. The lines
for the HO$_3$AlAl-$R$ and H$_3$O$_3$AlAl-$R$ 
surface energies calculated with respect to
$\mu_{\rm H}$ in equilibrium with H$_2$ and H$_2$O
cross at $\mu_{\rm O}$ equal to -2.7 eV at 0K. Below -2.7 eV, $\mu_{\rm H}$ is
determined by the equilibrium with H$_2$, 
and above with H$_2$O.
Below these lines either H$_2$ or H$_2$O would condense
on the surface.
It should be noted that the only independent variables in our system
is $\mu_{\rm O}$ in equilibrium with bulk Al$_{2}$O$_{3}$,
and $\mu_{\rm H}$. From these the dependent Al, O$_2$, H$_2$ and
H$_2$O chemical potentials follow.

The results are shown in Fig.~\ref{E-surf}. The Gibbs free energies
per surface area for various $(1 \times 1)$
geometries of the (0001) surface are displayed
as a function of $\mu_{\rm O}$. The dashed black 
vertical lines bracket the allowed range of $\mu_{\rm O}$. 
For the systems containing just aluminum and oxygen, there is clearly only
one overwhelmingly low energy surface stoichiometry: the AlO$_3$Al-$R$
structure. This is consistent with the previous theoretical results.
Our results show, however, that this stability extends
across the entire range of the oxygen chemical potential, even up to the
limit where an O$_2$ condensate would form on the surface. The 
empirical models explain this remarkable stability on the basis of
simple ionic considerations, as it is the only structure without a
significant surface
dipole. The actual situation is a bit more complicated, of course, and the
quantum mechanical calculations performed by us
and others\cite{GUO92,GOD94} indicate
that the surface aluminum
atom relaxes inward until it is practically coplanar with the oxygen layer,
and rehybridizes to an $sp^2$ orbital configuration which is dramatically
stabilized by autocompensation and bonding considerations.
That is, the structure relaxes and charge is transferred so that the $sp^2$ bonds
between the aluminum atoms and the three oxygens are filled, the aluminum
$3 p_z$ orbital perpendicular to the surface is empty, and there are no
partial occupancies of dangling bonds at the surface. In our calculations the
lowest empty surface state, which is primarily the Al $3 p_z$ with some
Al $3s$ and O $2p$ character, is 4.5 eV above the Fermi level. This is very high and
contributes to the relative chemical inertness of the relaxed Al-terminated surface.
Moving the topmost aluminum atom of the Al-terminated surface
from its position to one of the three non-bulk
positions indicated in Fig.~\ref{Al-surf}, increases the surface
energy dramatically from 133 meV/\AA$^2$ to 355, 282, and 228 meV/\AA$^2$
for the three sites, respectively.

The O$_3$-terminated surface has a very unfavorable 
energy, although this decreases
sharply as $\mu_{\rm O}$ increases. This stoichiometry has
the highest dipole at the surface, which is reflected in a very large 
work function of 9.66 eV, compared to 5.97 eV for the Al-terminated
surface above. In addition, there are partially filled open valencies
on the surface oxygens which are not relieved by relaxation.  Our results
indicate, however, that this high energy surface can become more stable
through oxygen evaporation. The O$_2$-terminated surface has a consistently lower
free energy than the O$_3$AlAl-$R$, and the O$_1$AlAl-$R$
surface is more stable than O$_2$AlAl-$R$,
except at the highest range of the oxygen chemical potential. 
At 1000K the metal-terminated surfaces become the lowest energy
surface under the most oxygen-deficient conditions.
This is consistent with what happens at annealing temperatures above
1250$^\circ$~C, where oxygen evaporation leaves behind a metallic 
aluminum overlayer \cite{VEH94}.
The figure does not convey the kinetic barriers required
to move from one surface stoichiometry to another, but the relative
thermodynamic stability is clear and is consistent with experimental
observations.

\begin{table}[t]
\caption{
Interlayer relaxations at the Al-terminated surface
in \% of  the corresponding bulk spacings.}
\label{table-Al}
\begin{tabular}{c|ccc|ccc}
          & \multicolumn{3}{c|}{Theory of Ref.}
          & \multicolumn{3}{c}{Experiment of Ref.}  \\
interlayer         &this             &\onlinecite{IM93}   &\onlinecite{CV98}
&\onlinecite{JA97} &\onlinecite{PG97}&\onlinecite{JT98}\\
\tableline
Al-O$_3$ \quad 1-2 &$-$86  &$-$86  &$-$87  &$-$63  &$-$51  & +30\\
O$_3$-Al \quad 2-3 &  +6   & +3    &  +3   &  --   & +16   & +6 \\
Al-Al    \quad 3-4 &$-$49  &$-$54  &$-$42  & --    &$-$29  &$-$55\\
Al-O$_3$ \quad 4-5 & +22   & +25   & +19   & --    & +20   & --\\
O$_3$-Al \quad 5-6 & +6    &  --   & +6    & --    & --    & --\\
\end{tabular}
\end{table}

The large surface dipole and dangling bonds of the O$_3$-terminated surface
can also be compensated by the addition of hydrogen to the surface. Adding
one hydrogen per unit cell to the surface \cite{hpos} dramatically 
lowers the free energy and the work function from
9.66 to 7.06 eV. This surface is much more stable than the
O$_3$-terminated surface at all oxygen partial pressures.
Saturating all
surface oxygens with hydrogen, indicated 
by the H$_3$O$_3$AlAl-$R$ line in Fig. 2, results in
the lowest free energy and greatest stability across the entire range of
physically realistic conditions. The oxygen open valencies are filled,
charge transfer occurs from the H to the topmost O layer,
and the work function is reduced to 3.44 eV, the lowest of all surfaces
examined.
The region of negative surface energy of the H$_3$O$_3$-terminated 
surface indicates that the hydroxylated surface is lower
in energy than bulk $\alpha$-Al$_{2}$O$_{3}$ plus free H$_2$O. This negative
surface energy reflects the strength of the H-OAl bond and is consistent
with the negative heat of the $\alpha$-Al$_{2}$O$_{3}$ + 3H$_2$O $ \rightarrow$
2Al(OH)$_3$ reaction. This reaction is exothermic by 0.19 eV per
Al$_{2}$O$_{3}$ formula unit 
relative to liquid water at 298.15K and 1 atm pressure \cite{NBS82}.

We find that in addition to explaining the relative stabilities of
the (0001) $(1 \times 1)$ surfaces observed experimentally, considering the presence
of hydrogen on the surface is essential to resolve the controversy
surrounding the surface relaxations. Our results are shown in
Table \ref{table-Al}. In the absence of hydrogen, our
full potential results find a large inward relaxation of 86\% for the first
layer, in very close agreement with previous pseudopotential
calculations\cite{IM93,CV98}, although there is less agreement for
subsequent layers.
When a hydrogen atom occupies site 1, however, the contraction between
the first Al and O layers is $\sim$ 69$\%$, in very close agreement
with the experimental result of Ahn and Rabalais \cite{JA97}. Several
other locations for the hydrogen atom were investigated, with similar
results. 

For the oxygen-only terminated surface, theoretical predictions have
indicated an inward relaxation of the top layer of 0.05\AA.
Yet the experimental results of
Toofan and Watson, in which one of the two domains observed was 
O-terminated, clearly showed an outward relaxation of 0.12\AA \cite{JT98}.
Placing a hydrogen at site 1, however, results in a predicted outward
relaxation of 0.11\AA, in excellent agreement with experiment.
Saturating the surface with hydrogen, the H$_3$O$_3$-structure,
increases the predicted relaxation to 0.19\AA.
Hence even though H was not detectable with the instrumentation used by
Toofan and Watson, our free energy calculations and predicted
relaxations provide strong evidence that H was most likely
present on the surface.

Real surfaces are usually much more
complicated and heterogeneous than the model $(1 \times 1)$ system
studied here, but encorporating the effects of chemical and thermal
equilibrium with the environment into the model provides a sound framework for
interpretation of experimental results on these real systems.


\begin{references}
\bibitem{hpos}  Several positions for a single hydrogen atom on
the O$_3$-terminated surface were evaluated. Three energetically
similar metastable positions were found, corresponding to the bulk
Al, and sites 1 and 2 shown in Fig.~\ref{Al-surf}.
\end{references}
\end{document}